\documentclass[pre,nofootinbib,twocolumn,showpacs,preprintnumbers,amsmath,amssymb,aps,10pt,superscriptaddress]{revtex4-1}
\pdfoutput=1
\usepackage{amsmath}
\usepackage{amssymb}
\usepackage{latexsym}
\usepackage{amsfonts}
\usepackage{graphicx}

\usepackage{color}
\usepackage[maxfloats=100]{morefloats}

\begin{document}

%\preprint{?}

\title{Frequency-Dependent Selection at Rough Expanding Fronts}

\author{Jan-Timm Kuhr}
%\author{Jan-Timm Kuhr and Holger Stark}
\email{jan-timm.kuhr@tu-berlin.de}
%\ead{jan-timm.kuhr@tu-berlin.de}
\affiliation{%
Institut f\"ur Theoretische Physik, Technische Universit\"at Berlin, Hardenbergstrasse 36, 10623 Berlin, Germany}
\author{Holger Stark}
\affiliation{Institut f\"ur Theoretische Physik, Technische Universit\"at Berlin, Hardenbergstrasse 36, 10623 Berlin, Germany}
%\address{Institut f\"ur Theoretische Physik, Technische Universit\"at Berlin, Hardenbergstrasse 36, 10623 Berlin, Germany}

\date{\today}% It is always \today, today,

\begin{abstract}
Microbial colonies are experimental model systems for studying
the colonization of new territory by biological species through range expansion.
We study a generalization of the two-species Eden model, which incorporates local frequency-dependent selection,
in order to analyze how social interactions between two species influence surface roughness of growing microbial colonies.
The model includes several classical scenarios from game theory. We then concentrate on an expanding public goods game,
where either cooperators or defectors take over the front depending on the system parameters. We analyze in detail the critical
behavior of the nonequilibrium phase transition between global cooperation and defection and thereby identify a new universality 
class of phase transitions dealing with absorbing states. At the transition, the number of boundaries separating sectors 
decays with a novel power law in time and their superdiffusive motion crosses over from Eden scaling to a nearly ballistic regime. 
In parallel, the width of the front initially obeys Eden roughening and, at later times, passes over to selective roughening.
%If colonies consist of more than a single species a multitude of patterns is observed, especially if species influence each 
%other's reproduction. Lattice models from non-equilibrium statistical mechanics reproduce many phenomena, like the formation 
%of single-species sectors and the roughness of the growing colony's front. Here, we set up a novel generalization of the classical %Eden model to study multi-species colony growth.
%It combines the dynamics of a rough, expanding front with local interactions known from game theory. 
%Our model covers competition between species reproducing at different rates, but more importantly a variety of classical 
%scenarios form game theory. For a public-good-interaction, we find a transition between global cooperation and defection, in the 
%form of a nonequilibrium phase transition. While the transition constitutes a new universality class of phase transitions dealing 
%with absorbing states, it also has impact on observable patterns of the colony. In particular we encounter a new surface 
%roughening regime, which also results in a qualitative change in the superdiffusice dynamics of sector boundaries.
\end{abstract}

%02.50.Le	Statistical Physics - Probability theory, stochastic processes, and statistics -  Decision theory and game theory
%68.35.Ct		Surfaces and interfaces; thin films and nanosystems structure and nonelectronic properties - Solid surfaces and solid-solid interfaces: structure and energetics - Interface structure and roughness
%68.35.Ct		Surfaces and interfaces; thin films and nanosystems structure and nonelectronic properties - Solid surfaces and solid-solid interfaces: structure and energetics - Phase transitions and critical phenomena
%87.18.Hf		Biological and medical physics - Biological complexity - Spatiotemporal pattern formation incellular populations
%87.23.-n 	Biological and medical physics - Ecology and evolution
%\noindent{\it Keywords}: non-equilibrium phase transition, surface growth, evolutionary game theory, Eden model, evolution, range expansion
%\submitto{\NJP}

\maketitle

\section{Introduction}
Living species are usually confined to their territory, a spatial region defined by geographical borders, climate, or other environmental constraints.
Uninhabited regions are colonized through range expansion,
where individuals reproduce and disperse at the front 
%(i.e., the part of the population at the very border to the unoccupied region)
of their territory~\cite{Murray:2007ux}.
This process is seen in biological invasions~\cite{Griffiths:1991tea}, as a result of shifting climate zones~\cite{Loarie:2009gx,Chen:2011fl,Peacock:2015ch}, during colonizations in our own species' history~\cite{Stringer:2003id,Liu:2006gv,Moreau:2011iv}, tumor growth~\cite{Murray:2007ux,Bru:2003fa,Xavier:2011dv}, and biofilm growth~\cite{HallStoodley:2004cv,Nadell:2009de,Mitri:2011fs}. 
Evidently, expansions occur on very different spatial (micrometers to  $10^7$ meters) and temporal (hours to millennia) scales.

In this article we aim to characterize range expansion under the influence of short-range ``social interactions'' of individuals at the front.
Such interactions are present if success in reproduction depends on the presence of nearby individuals of the own and/or other species.
Here, we set up a model for the expanding front based on evolutionary game theory~\cite{Hofbauer:1998wn,Szabo:2007eq,Frey:2010iz} and investigate its roughening dynamics for two interacting species.
Besides exploring an interesting non-equilibrium growth process, we hope to contribute to interpreting experiments on range expansion in multi-species colonies of simple organisms.

In experiments, microbial growth is excellently suited to study range expansion and other processes in population dynamics and evolution such as spatial spread of infections and  adaptation to an environment (see for example Ref.~\cite{Buckling:2009je}). 
Microbes reproduce fast, their environment and genotype can be controlled, and experimental conditions are easily reproducible.
Grown in a Petri dish, the spatial patterns of single-species microbial colonies have long been a rich field of study~\cite{Shapiro:1995id,Jacob:1999bn,Golding:1999wa,Matsushita:1999vr,Matsuyama:2001vy}.
The observed patterns crucially depend on motility, availability of nutrients, and the growth medium, to name but a few.
However, even under conditions of negligible motility and abundant nutrients a colony's front is rough and has interesting statistical properties~\cite{Vicsek:1990wy,Wakita:1997do,Huergo:2010bb}.

Multi-species colonies are composed of more than one species and show additional intriguing features, even if the species are identical except for a marker~\cite{Hallatschek:2007gv}.
During reproduction they keep their marker but compete with other species for space at the front. 
Thereby, \emph{sectors of single species} form, which are separated by \emph{boundaries}. 
Their statistical and dynamic properties are determined by the evolving roughness of the expanding front~\cite{Derrida:1991tv,Saito:1995wl}.
%When boundaries meet, the sector in between looses contact to the front and the sectors on either side merge. As clusters continue to merge the number of surviving sectors is diminished~\cite{Hallatschek:2007gv,Ali:2010do}.

Usually, when two or more species of microbes live in a common environment, they influence each other during reproduction.
In particular, reproductive success of any species, also called its fitness, depends on the population sizes of all the species.
This constitutes ``social interactions'' between the species commonly referred to as \emph{frequency-dependent selection}. 
Research in the field has initiated a wealth of fascinating experiments~\cite{Crespi:2001fy,Velicer:2002cs,Griffin:2004kn,Kreft:2004uz,West:2007kq,Hibbing:2010jw,Xavier:2011dv,Mitri:2013gx} either in well-mixed liquid culture without any spatial order~\cite{Griffin:2004kn,Gore:2009jk} or in the Petri dish, where the populations are spatially structured~\cite{Muller:2014ev,VanDyken:2013tf}.
Many of the experimental observations can be discussed within the framework of evolutionary game theory~\cite{Hofbauer:1998wn,Szabo:2007eq,Frey:2010iz}.
For example, light has been shed on a long-standing theoretical question in 
evolution~\cite{Darwin:1859um,Hofbauer:1998wn,Axelrod:2009vw,Szabo:2007eq}: 
Why do individuals cooperate if non-cooperators can exploit them?
%How can cooperation persist, given that it can easily be exploited by individuals which do not cooperate?
Literature emphasizes the importance of a population to be structured in groups~\cite{Rainey:2003iz,Chuang:2009hu,Nowak:1994vj,Ohtsuki:2006cq,Szabo:2007eq,Fu:2010kp}, for example, by spatial distance.
While within a single group cooperators are always inferior to non-cooperating defectors, 
if the latter interact only with their neighborhood,
distant large groups of cooperators will ultimately outcompete defectors.
%If groups compete with each other, however, cooperators may prevail, as groups with a large fraction of cooperators outcompete groups of defectors.
Some models also stress the central role of demographic fluctuations and of populations growing in size~\cite{Melbinger:2010th, Cremer:2012vk}.
Both, experiments and theory, explain the advantage of cooperators during colony growth by their ability to locally advance faster~\cite{Xavier:2007gp,Nadell:2010fe,VanDyken:2013tf}, vividly termed ``survival of the fastest''~\cite{VanDyken:2013tf}.

Cooperation between between nearby cells is often mediated by some biochemical compound (a \emph{public good}) which the microbes release into the extracellular environment.
This compound then promotes reproduction of neighboring cells.
In general, a released substance may act beneficial or detrimental to other individuals, also depending on their species, and implies some cost to the producer. 
Examples include secretion of digestive invertase to break down 
sucrose~\cite{Greig:2004hb,Gore:2009jk,HKoschwanez:2011bc,SenDatta:2013gj},
siderophores to scavenge iron from the environment~\cite{Pattus:2000vh,Ratledge:2000iv,Griffin:2004kn}, 
polymers which support biofilms~\cite{Rainey:2003iz,Nadell:2011jf,vanGestel:2014dt}, 
release of toxins~\cite{Riley:1999ws,Cornforth:2013fw} 
(sometimes through lysis~\cite{Lee:2001ii,Weber:2014hz}), 
surfactants which facilitate swarming~\cite{Xavier:2011js}, and the exchange of amino acids~\cite{Muller:2014ev}.

This plethora of 
%\jtk{intercellular interactions}
%\jtk{reciprocal actions between cells} 
%\hs{biochemical compounds acting as public goods}
biochemical compounds, released by cells and affecting nearby cells,
implies a wealth of specific features, which certainly are not covered by a single model.
However, since 
%these 
%interactions between individuals are usually 
%\hs{public goods}
the released biomolecules 
usually mediate short-range interactions between individuals,
%short-ranged, 
properties on large scales should be independent of microscopic details.
Hence, we formulate a simple model which captures the essence of an interaction while ignoring complicated details.
The classical Eden model~\cite{Eden:1960vd, Barabasi:1995vz}, a simple growth process on a lattice, has been used successfully to mimic growing cell colonies..
It generates a cluster (the colony), the surface of which exhibits scaling properties also found for expanding fronts of 
microbial colonies~\cite{Plischke:1985td,Jullien:1985th}. 
Extended to two identically growing but still distinguishable species, it generates sectors occupied by a single species only~\cite{Saito:1995wl,Ali:2010do}.
Indeed, this behavior is found for two-species microbial colonies~\cite{Hallatschek:2007gv}. 
Moreover, boundaries between sectors move superdiffusively as in the experiments.

In this article we explore a generalization of the two-species Eden model, which incorporates local frequency-dependent selection.
We thereby aim to analyze how social interactions influence surface roughness of growing microbial colonies.
We set up an expanding public goods game, where either cooperators or defectors take over the front depending on the system
parameters~\cite{Hofbauer:1998wn,Szabo:2007eq,Frey:2010iz}. Right at the transition the front displays critical behavior,
which we analyze in detail. 
In particular, we establish that our model belongs to a new universality class of phase transitions dealing with absorbing states.
At the transition, the number of boundaries separating sectors decays with a novel power law in time and their
superdiffusive motion crosses over from Eden scaling to a nearly ballistic regime. In parallel, the width of the front 
initially obeys Eden roughening and, at later times, passes over to what we call selective roughening.

The remainder of this article is organized as follows. To analyze multi-species microbial colony growth,
we introduce the Eden model with frequency-dependent selection in Sec.~\ref{sec:model} and analyze its phenomenology in Sec.~\ref{sec:phen}.
We then concentrate on the expanding public goods game with its social dilemma in Sec.~\ref{sec:critical_behavior}
and analyze the critical behavior at the transition between long-term cooperation and long-term defection by applying 
statistical analysis.
Finally, we discuss and summarize our findings in Sec.~\ref{sec:discussion}.

\section{Eden Model with Frequency-Dependent Selection \label{sec:model}} 

\begin{figure}%
\centering%
%\begin{indented}\item[]
\includegraphics[width=\linewidth]{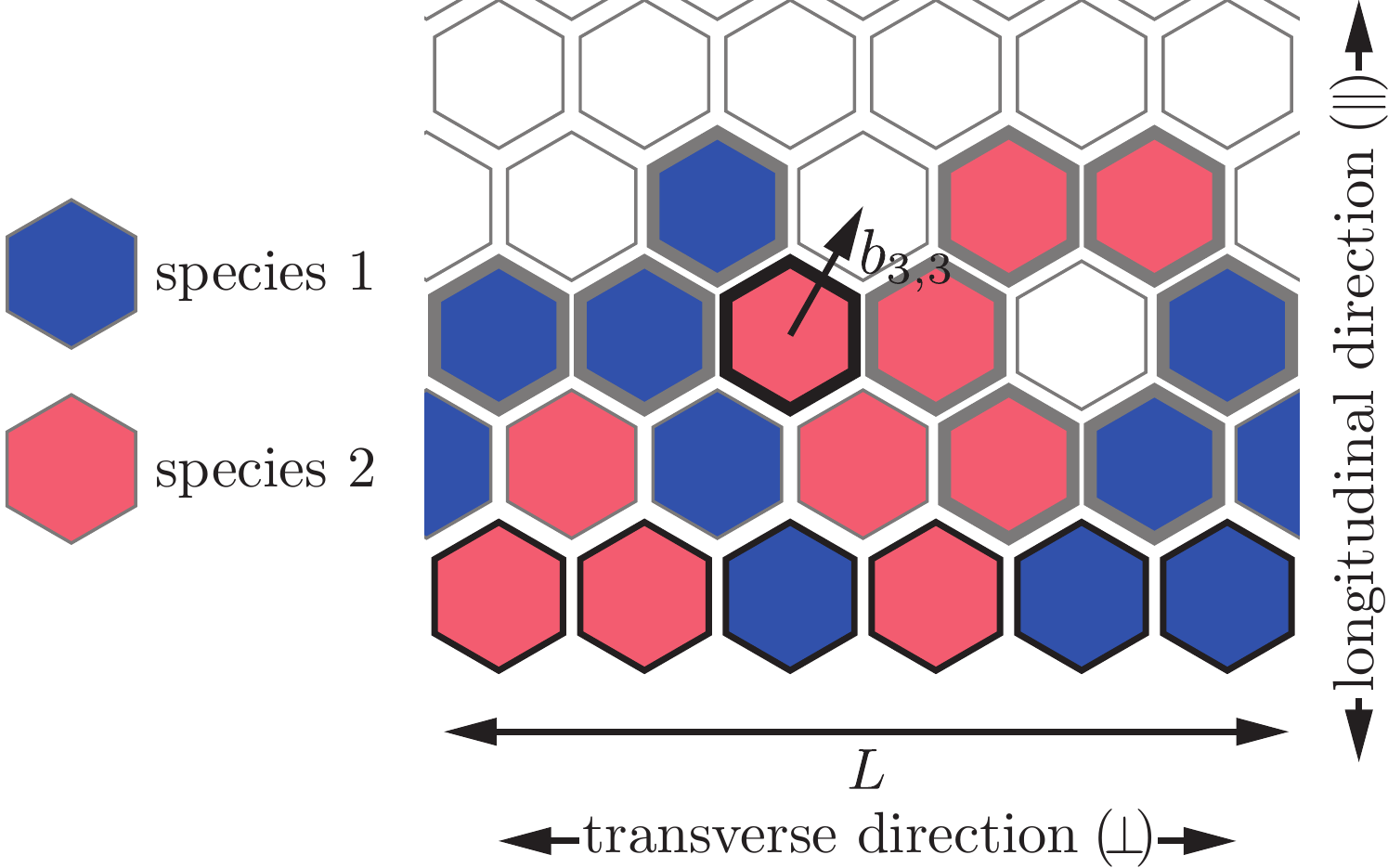}%
%\end{indented}
\caption{\label{fig:model} 
Two-species Eden model with frequency-dependent selection on a hexagonal lattice. 
The bacterial colony grows from the bottom line (lattice sites with narrow black edge) of length $L$, where individuals of species 1 (blue) and species 2 (red)
occupy the lattice sites.
The colony expands into the empty, infinitely-extended half-space.
Individuals capable of reproduction (indicated by bold edges), have at least one empty lattice site as a nearest neighbor. 
In a reproduction event, one of these empty neighboring sites $(i,j)$ is chosen with equal probability 
and the reproducing individual changes the corresponding state $s_{i,j}$ to its own state 1 or 2.
Each individual has its own reproduction rate $b_{i,j}$ given in Eq.~\eqref{eq:br_lattice}.
%which depends i) on its species and ii) on the number of individuals of species 1 and 2 in its immediate neighborhood. 
For example, the reproduction rate of the individual at site $(3, 3)$ (bold black edges) is 
$b_{3,3} = b_2^0 + 3\mathcal{T}+2\mathcal{P}$.
Along the transverse direction periodic boundary conditions apply.}
\end{figure}

In this work we employ a lattice model (see Fig.~\ref{fig:model}) to analyze range expansion at rough fronts under the influence of frequency-dependent selection. 
We set up a cellular automaton on a two-dimensional hexagonal%
\footnote{On a square lattice it is impossible to enclose a cluster $A$ within a cluster $B$, which only contains nearest neighbor sites of cluster A. On a hexagonal lattice this is possible.}
lattice of transverse extension $L$ and an infinite longitudinal extension.
Periodic boundary conditions are applied in the transverse direction.
The state $\{s\}$ of the system at time $t$ is specified by the state variables $s_{i,j}$ of lattice sites $(i,j)$. 
Consider a system with two species (extension to more species is straightforward). 
For any time $t$, any site  $(i,j)$ is either empty ($s_{i,j} = 0$) or occupied by an individual of either species 1 ($s_{i,j} = 1$) or 2 ($s_{i,j} = 2$).
All individuals which have at least one free nearest neighbor site can reproduce.
To perform a reproduction step, we choose one of these fecund individuals with a probability proportional to its reproduction rate (see below) 
and a new individual of the same species is placed with equal probability on one of the free neighboring sites.

In contrast to the Eden model~\cite{Eden:1960vd} and some of its two-species generalizations~\cite{Saito:1995wl,Kuhr:2011cq}, 
reproduction rates in our model depend on the states of the nearest-neighbor sites.
Let $n_1$ and $n_2$ denote the number of nearest neighbors of species 1 and 2, respectively, then the reproduction rate of an individual at lattice site $(i,j)$ is 
\begin{eqnarray}
b_{i,j} = 
\begin{cases}
0 & \text{if } s_{i,j} \text{ has no free neighbors}, \\
b^0_1 + n_1 \mathcal{R}+ n_2 \mathcal{S} & \text{if } s_{i,j} = 1 ,\\
b^0_2 + n_1 \mathcal{T} + n_2 \mathcal{P} & \text{if } s_{i,j} = 2 .
\end{cases} \label{eq:br_lattice}
\end{eqnarray}
%
%\begin{eqnarray}
%b_{i,j} = 
%\left\{
%\begin{array}{ll}
%0 & \text{if } s_{i,j} \text{ has no free neighbors}, \\
%b^0_1 + n_1 \mathcal{R}+ n_2 \mathcal{S} & \text{if } s_{i,j} = 1 ,\\
%b^0_2 + n_1 \mathcal{T} + n_2 \mathcal{P} & \text{if } s_{i,j} = 2 .
%\end{array}
%\right.
%\label{eq:br_lattice}
%\end{eqnarray}
%
Here, $b^0_1$ and $b^0_2$ are the respective contributions to the reproduction rates of species 1 and 2,  which are independent of the states of their nearest neighbors. 
Frequency-dependent selection is introduced through the parameters $\mathcal{R}$, $\mathcal{S}$, $\mathcal{T}\!$, and $\mathcal{P}$.

With the reproduction rates $b_{i,j}$ we implement a random sequential update of the system
using a simplified version of the Gillespie algorithm~\cite{Gillespie:1976vd}.
The overall reproduction rate of the population is 
$b_{tot} := \sum_{i,j} b_{i,j}$ 
and an individual at site $(i, j)$ is selected to reproduce with probability $b_{i,j}/b_{tot}$.
We then choose one of the empty nearest-neighbor sites of the reproducing individual at random and place there a new individual of the same species. 
This implies that there are no mutations. 
Since the mean time until the next reproduction event is $b_{tot}^{-1}$, we update time by $t \rightarrow t + b_{tot}^{-1}$ after each reproduction event.
We assume that individuals do not die and that they are immobile.
Therefore, any site with $s_{i,j} \neq 0$ remains in its specific state indefinitely.
As initial condition we occupy all sites of an initial line randomly, but in equal parts, with species 1 and 2, if not stated otherwise.

The formulated model  generalizes version C of the Eden model, introduced by Jullien and Botet~\cite{Jullien:1985wv}, to a two-species system.
We already applied a similar model to range expansion 
without frequency-dependent selection but included the possibility of mutations~\cite{Kuhr:2011cq}.
%Version C is biologically more realistic compared to A and B, as it focuses on occupied sites, i.e., individuals, rather than on empty sites (version A) or bonds between adjacent occupied and empty sites (version B). 
If $\mathcal{R}$, $\mathcal{S}$, $\mathcal{T}\!$, and $\mathcal{P}$ are zero, our model reduces to that of Saito and M\"uller-Krumbhaar~\cite{Saito:1995wl}, however they used a square lattice.
Since diffusion is not included in the model, configurations and patterns behind the front are frozen.
This corresponds to observations in microbial experiments on range expansion~\cite{Hallatschek:2007gv,Weber:2014hz}.

%\footnote{The specific letters come from the classic game ``prisoner's dilemma'' and stand for ``reward (for cooperation)``, ``sucker's payoff``, ``temptation (to defect)``, and ``punishment``.}
In game theory the parameters $\mathcal{R}$, $\mathcal{S}$, $\mathcal{T}\!$, and $\mathcal{P}$ from Eq.~\eqref{eq:br_lattice} 
define the payoff matrix of a 
%symmetric two-player, 
two-strategy game~\cite{Hofbauer:1998wn,Szabo:2007eq,Frey:2010iz}. 
%\begin{center}%
%\begin{tabular}{cc|cc}%
%\multicolumn{2}{c}{}&\multicolumn{2}{c}{oponent's}\\
%\multicolumn{2}{c}{}&\multicolumn{2}{c}{strategy}\\
%&&$C$&$D$\\ \cline{2-4}%
%own &$C$&$\mathcal{R}$&$\mathcal{S}$\\
%strategy&$D$&$\mathcal{T}\!$&$\mathcal{P}$
%\end{tabular}%
%\end{center}%
%Here the entries give the payoff which depend on one's own and the opponents strategy, where the strategies are typically denoted by ``cooperate'' ($C$) and ``defect'' ($D$). 
%In our case each individual interacts with all its nearest neighbors, which creates, in general, a game with more than two players.
Different scenarios, some well known in game theory, are implemented if we set these parameters accordingly.

\section{Phenomenology\label{sec:phen}}
\begin{figure}%
\centering%
\includegraphics[width=\linewidth]{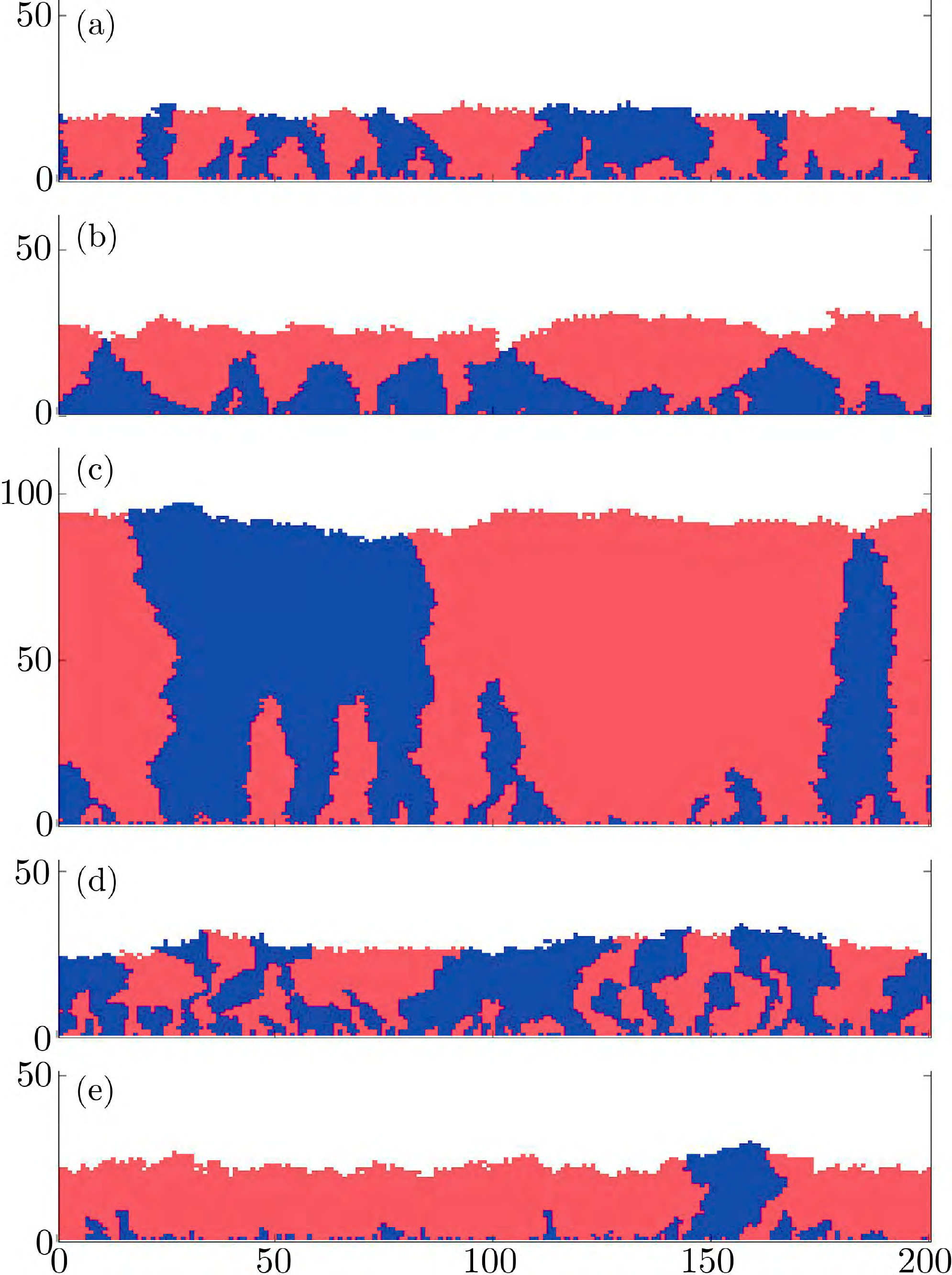}%
%\begin{indented}\item[]
%\includegraphics[width=\linewidth]{figure2.eps}%{pics/typicalsettings.pdf}%
%\end{indented}
\caption{\label{fig:settings} 
Growth patterns of our model for different parameters, which correspond to typical settings.
Time is always $t=15$ and lattice size is $L=200$. 
Cooperators are depicted in blue and defectors in red.
If not stated otherwise $b_2^0 = b_2^0 = 1$, $\mathcal{R} = \mathcal{P} = \mathcal{S} = \mathcal{T} = 0$, and the initial ratio of both species is 1:1. 
(a)~neutral growth,
%. Coarsening of sectors by merging of meandering boundaries at the non-flat front can be observed. 
(b)~selective advantage for species 2 ($b_2^0 = 1.5$), which occupies 10\% of initial sites, 
%Boundaries are biased, widening sectors of species 2, which is about to fix at the front, thereby taking the system into an absorbing state. The front expands faster where it is composed of species 2, which results in enhanced front roughness. 
(c)~coordination game ($\mathcal{R} = \mathcal{P} = 1$), 
%Expansion advances fast, as sectors coarsen and most individuals have only neighbors of their own kind. Dents tend to be close to sector boundaries. 
(d)~snowdrift game ($\mathcal{S} = \mathcal{T} = 1$), 
%Boundaries annihilate less frequent as compared to (c), are strongly twisted, and tend to be near bulges of the front. 
and
(e)~public goods game ($\mathcal{R} = 0.1,  \mathcal{T} =1.1$).
% with cooperators in blue and defectors in red. 
%Species 2 (defectors) rapidly take over wide sections of the front, since initially boundaries are numerous and defectors benefit from neighbors of species 1 (cooperators). The latter only stand a chance where, by chance, they form a sufficiently large sector, which can then outgrow defectors trailing its boundaries.
}
\end{figure}

We now describe some generic examples of our model for growing microbial colonies (see Fig.~\ref{fig:settings}), 
which emerge for typical parameter settings, 
and discuss their characteristic features.
We then concentrate on so-called social dilemmas, where one species (defectors) exploits the other one (cooperators).
Due to the spatial extent of our system, cooperators are able to outcompete defectors in a defined parameter region. This is
in contrast to a single group, where all members interact with each other and, therefore, cooperators are always inferior to 
defectors.

In the simplest case selection is frequency-independent, $\mathcal{R} = \mathcal{S} = \mathcal{T} = \mathcal{P} = 0$, and 
both species reproduce with the same rate, $b^1_0 = b^2_0$ [see Fig.~\ref{fig:settings}(a)].
For this ``neutral'' setting we observe roughening of the front typical for the Eden model~\cite{Eden:1960vd, Barabasi:1995vz}.
Simultaneously, sectors composed of a single species merge and thereby coarsen~\cite{Saito:1995wl,Ali:2010do}.
This inherent process happens according to the following scenario. 
If the tips of two advancing boundary lines meet, they annihilate%
%\footnote{In models with more than two species it is possible that two boundaries merge into a single one.} 
and the enclosed sector loses contact to the front.
Consequently, the number of boundaries and sectors can only decrease. 
Sectors repeatedly coarsen as they merge in these events.
When all boundaries have vanished, the front ``has fixed'' to a single species keeps on expanding.
In finite systems, $L < \infty$, fixation to a single species always occurs since in our stochastic model there is always a finite rate at which boundaries annihilate.
%(if the entries of the payoff matrix are finite at least)
Hence, two absorbing states exist. 
Eventually, the expanding front will fix either to species 1 or species 2.

In Fig.~\ref{fig:settings}(b), species 2 has a larger reproduction rate, $b^2_0 > b^1_0$, and therefore a constant selective advantage. 
As the front expands faster at locations where it is composed of species 2, the roughness of the front increases.
Indentations of the front usually are caused by sectors of species 1, whereas species 2 creates bulges.
Furthermore, boundaries are biased such that sectors of species 2 widen while sectors of species 1 shrink laterally.
Hence, sectors of species 2 merge and coarsen quickly.
Eventually, the expanding front will fix to species 2, which has almost happened in Fig.~\ref{fig:settings}(b).

If the reproduction rates depend on the state of nearest neighbors (frequency-dependent selection), new patterns arise.
In this article we are mainly interested in frequency-dependent selection and therefore set $b^0_1 = b^0_2 = 1$ from here on.
We now discuss some interesting cases, see Fig.~\ref{fig:settings}(c)--(d), which correspond to well-known settings in game theory~\cite{Hofbauer:1998wn,Szabo:2007eq,Frey:2010iz}.

In coordination games
($\mathcal{R} > \mathcal{T} \text{ and } \mathcal{P} > \mathcal{S}$), see Fig.~\ref{fig:settings}(c), the front expands 
slower near boundaries than in the centers of sectors. 
Therefore, indentations in the front are typically found where boundaries currently are or recently have been.
After sectors have coarsened for some time, most individuals are located inside sectors.
Therefore, most of them only have neighbors of their own kind,
which raises the average reproduction rate and the overall front advances faster.

In snowdrift games ($\mathcal{R} < \mathcal{T} \text{ and } \mathcal{P} < \mathcal{S}$), see Fig.~\ref{fig:settings}(d), the front expands faster near boundaries. 
They annihilate less frequent as compared to Fig.~\ref{fig:settings}(c) since narrow sectors grow faster. 
Boundaries are also strongly twisted and associated with bulges of the front.

In this article we are mostly interested in social dilemmas where one species (called cooperators) raises the reproduction rate of all neighbors regardless of their species.
The increased reproduction rate is called a \emph{public good} in game theory, since it is of benefit to all nearby individuals, but it also costs resources.
In contrast, the other species (called defectors) takes advantage of the public good for its own reproduction, but does not 
contribute to the reproduction of its neighbors in the same way.
Defectors save resources for their own reproduction and therefore have an advantage. 
In this scenario defectors
do not at all contribute to reproduction of their neighbors and, therefore, we set $\mathcal{S} = \mathcal{P} = 0$ and just vary $\mathcal{R}$ and $\mathcal{T}\!$.
According to Eq.~\eqref{eq:br_lattice}, these two parameters increase the respective reproduction rates of cooperators and defectors
if they have cooperating neighbors.
In Fig.~\ref{fig:settings}(e) we present a setting, where species 2 (defectors) rapidly takes over large parts of the front.
Defectors benefit from the initially large number of boundaries, where they take advantage of nearby cooperators, and 
conquer most of the front.
Only cooperators, living in sufficiently large sectors, can keep up with the front during this early period and may then take over the front, depending on the parameter values.
%Only sufficiently large cooperator sectors can keep up with the front during this early period. 

\begin{figure}%
\centering%
\includegraphics[width=\linewidth]{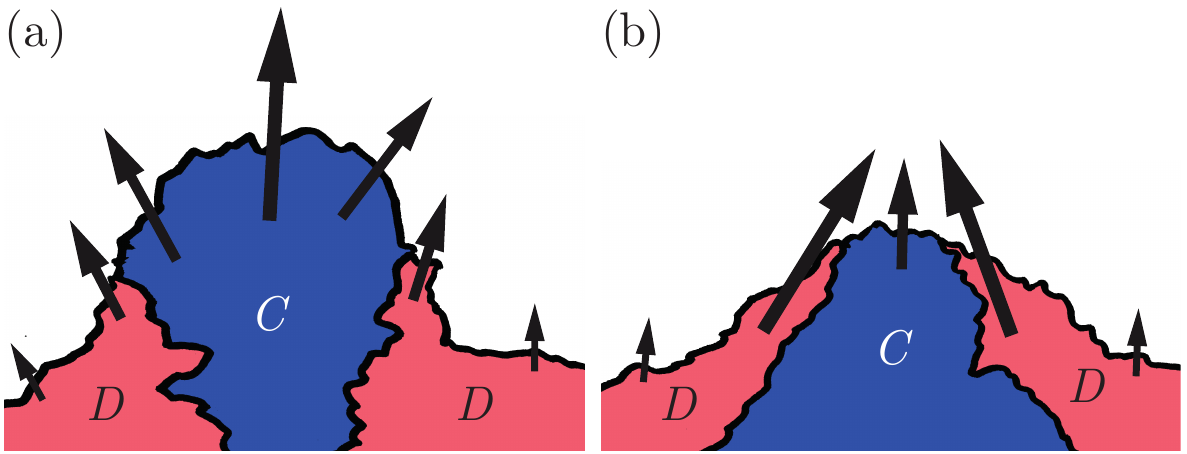}%
%\begin{indented}\item[]
%\includegraphics[width=\linewidth]{figure3.eps}%{pics/pgg_scematic_scenarios.pdf}%
%\end{indented}
\caption{\label{fig:pgg_scematic_scenarios}
Schematics of possible scenarios in an expanding public goods game.
Depending on the parameters $\mathcal{R}$, $\mathcal{S}$, $\mathcal{T}\!$, and $\mathcal{P}$ of Eq.~\eqref{eq:br_lattice}, cooperators ($C$, blue) and defectors ($D$, red) advance with different speeds, as indicated by arrows. 
(a) Cooperators outrun the trailing defectors.
From an advanced position along the front cooperators can then expand laterally and take over the front. 
(b) A thin layer of defectors keeps up with the cooperators' sector and, eventually, completely covers the cooperators.
}%
\end{figure}%

%The dilemma arises if defectors fix at the front, which then advances slower than an all-cooperator front would. 
%This implies a hierarchy of growth rates at the front:\\
%Defectors at boundaries reproduce fastest, followed by cooperators in the center of their sectors. Next fastest grow cooperators at boundaries. Defectors within sectors reproduce slowest. As a result, the front is advanced for cooperator sectors and falls back where defectors dominate. 
%At boundaries defectors tend to sideways ``overgrow'' cooperators.
In a situation like this, it is not \emph{a priori} clear if the front eventually fixes either to cooperators or to defectors.
Depending on the values of $\mathcal{R}$ and $\mathcal{T}\!$, cooperators can either outrun defectors and, 
from their advanced position at the front, overgrow their competitors laterally, see Fig.~\ref{fig:pgg_scematic_scenarios}(a).
Or, defectors cover cooperators with a thin layer and thereby take over the front, see Fig.~\ref{fig:pgg_scematic_scenarios}(b).
Close to the transition between both scenarios, the front displays increasing roughness since both species are able to take over while their fronts grow with different speeds.
To characterize this transition quantitatively, we performed extensive simulations and applied methods from surface roughening~\cite{Barabasi:1995vz,Krug:1992wu,HalpinHealy:1995wb,Odor:2004wm} and the theory of phase transitions dealing with absorbing states~\cite{Henkel:2008wn,Odor:2004wm}.

\section{Expanding Public Goods Game: Critical Behavior\label{sec:critical_behavior}}

In this section we quantitatively analyze the transition between long-term cooperation and long-term defection for an expanding public goods game. 
As the transition is approached, several observables show critical scaling~\cite{Henkel:2008wn,Odor:2004wm}.
Following our earlier work~\cite{Kuhr:2011cq}, we perform finite-size scaling to localize the transition. 
Furthermore, we determine critical exponents and thereby establish a new universality class for the transition between the two adsorbing states.
In the vicinity of the transition we also study the dynamics of the sector boundaries including the decline of their mean number during coarsening and their superdiffusive motion as well as the roughening of the expanding front.

\subsection{Finite-size scaling and phase diagram \label{sec:phase_trans}}

As boundaries merge, sectors coarsen and the system progresses towards one of the two absorbing states. 
At finite system sizes $L$ this is a stochastic process and both adsorbing states are reached with a certain probability.
However, in the thermodynamic limit, $L \to \infty$, the magnitude of fluctuations relative to the mean value goes to zero and one of the absorbing states is reached with certainty.
We now use the method of finite-size scaling to determine the transition point between both states~\cite{Henkel:2008wn}.

\begin{figure}%
\centering%
%\begin{indented}\item[]
\includegraphics[width=\linewidth]{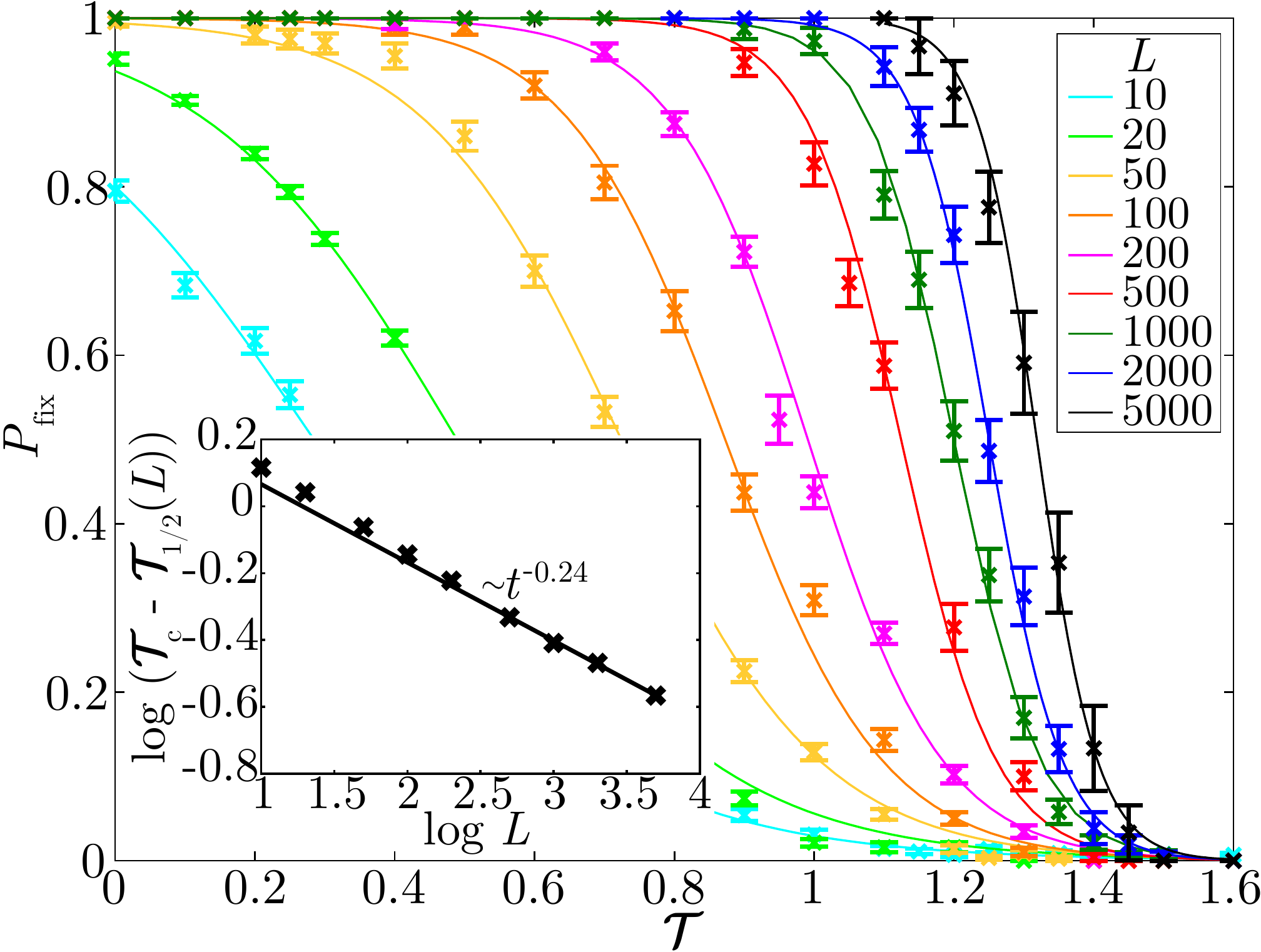}%
%\end{indented}
\caption{\label{fig:pfix_C}
Probability $P_\mathrm{fix}$ of the front to fix to cooperators plotted versus $\mathcal{T}\!$ for several system sizes $L$ at $\mathcal{R} = 0.1$.
%For larger $L$ the transition between cooperator fixation and defector fixation becomes steeper and shifts to larger values of $\mathcal{T}\!$. 
Error bars give the standard error of the mean for each data point.
Lines are fits of the data points to the Fermi function,
$1/\exp(c(\mathcal{T} - \mathcal{T}_{1/2}) + 1)$,
where $\mathcal{T}_{1/2}$ and $c$ are fit parameters.
Inset: The transition point $\mathcal{T}_{1/2}(L)$ relative to the fitted critical value $\mathcal{T}_c \approx1.58$ (black crosses)
follows a power law in $L \to \infty$: $\mathcal{T}_c - \mathcal{T}_{1/2}(L) = (A/L)^{1/4.2}$ (black line).
%(black crosses) approaches the critical value $\mathcal{T}_c$ for $L \to \infty$. The black line is a fit 
% \jtk{$\mathcal{T}_c - \mathcal{T}_{1/2}(L)$ approaches a power law (bold black line) for $L \to \infty$.
%Hence, we found an asymptotic scaling function for $\mathcal{T}_{1/2}(L)$, as given in Eq.~\eqref{eq:Tcrit}.}
%The transition point $\mathcal{T}_{1/2}(L)$ (black crosses) approaches the critical value $\mathcal{T}_c$ for $L \to \infty$. 
%The black line is a fit to the scaling function $\mathcal{T}_{1/2}(L) = \mathcal{T}_c - (A/L)^{1/\nu_\perp}$.
}
\end{figure}

In Figure~\ref{fig:pfix_C} we present the probability $P_\mathrm{fix}$ that the front fixes to cooperators and plot it versus $\mathcal{T}\!$ for several lattice sizes $L$ at $\mathcal{R} = 0.1$.
We distinguish two regimes: one where cooperators dominate ($P_\mathrm{fix}(\mathcal{T},L) > \frac12$) and one where
defectors take over.
We locate the transition point at $\mathcal{T}_{1/2}(L)$ by $P_\mathrm{fix}(\mathcal{T}_{1/2}(L),L) = \frac12$.
As $L \to \infty$, $P_\mathrm{fix}$ converges to a step function, since in infinite systems the absorbing states are reached with certainty.
The step is positioned at $\mathcal{T}_c := \lim_{L \to \infty}\mathcal{T}_{1/2}(L)$.
From the theory of critical scaling applied to absorbing states, we expect that close to the critical point $\mathcal{T}_c$ the states of the lattice sites are correlated on the transverse distance $\xi_\perp$.
Approaching $\mathcal{T}_c$, $\xi_\perp$ diverges as $\left|\Delta\right|^{-\nu_\perp}\!$, where $\Delta := \mathcal{T}_c - \mathcal{T}$ is the distance to the critical point and $\nu_\perp$ is a \emph{critical exponent}. 
For finite systems, an absorbing state is reached if 
\begin{align}
L \approx \xi_\perp \sim \left|\Delta\right|^{-\nu_\perp} \, .
\label{eq:critscalperp}
\end{align}
The transition occurs at $\mathcal{T}_{1/2}(L)$ and rearranging Eq.~\eqref{eq:critscalperp}, we obtain
\begin{align}
\mathcal{T}_{1/2}(L) \approx \mathcal{T}_c - (A/L)^{1/\nu_\perp} \; . \label{eq:Tcrit}
\end{align}
The characteristic length $A$ is related to the microscopic length scale, which here is the lattice constant,
and details of our model. 
It is not important to the following analysis.
The inset of Fig.~\ref{fig:pfix_C} shows the best fit of our data to Eq.~\eqref{eq:Tcrit}, which yields the critical exponent $\nu_\perp \approx 4.2$ and the critical point $\mathcal{T}_c \approx 1.58$ at $\mathcal{R} = 0.1$.

\begin{figure}%
\centering%
%\begin{indented}\item[]
\includegraphics[width=\linewidth]{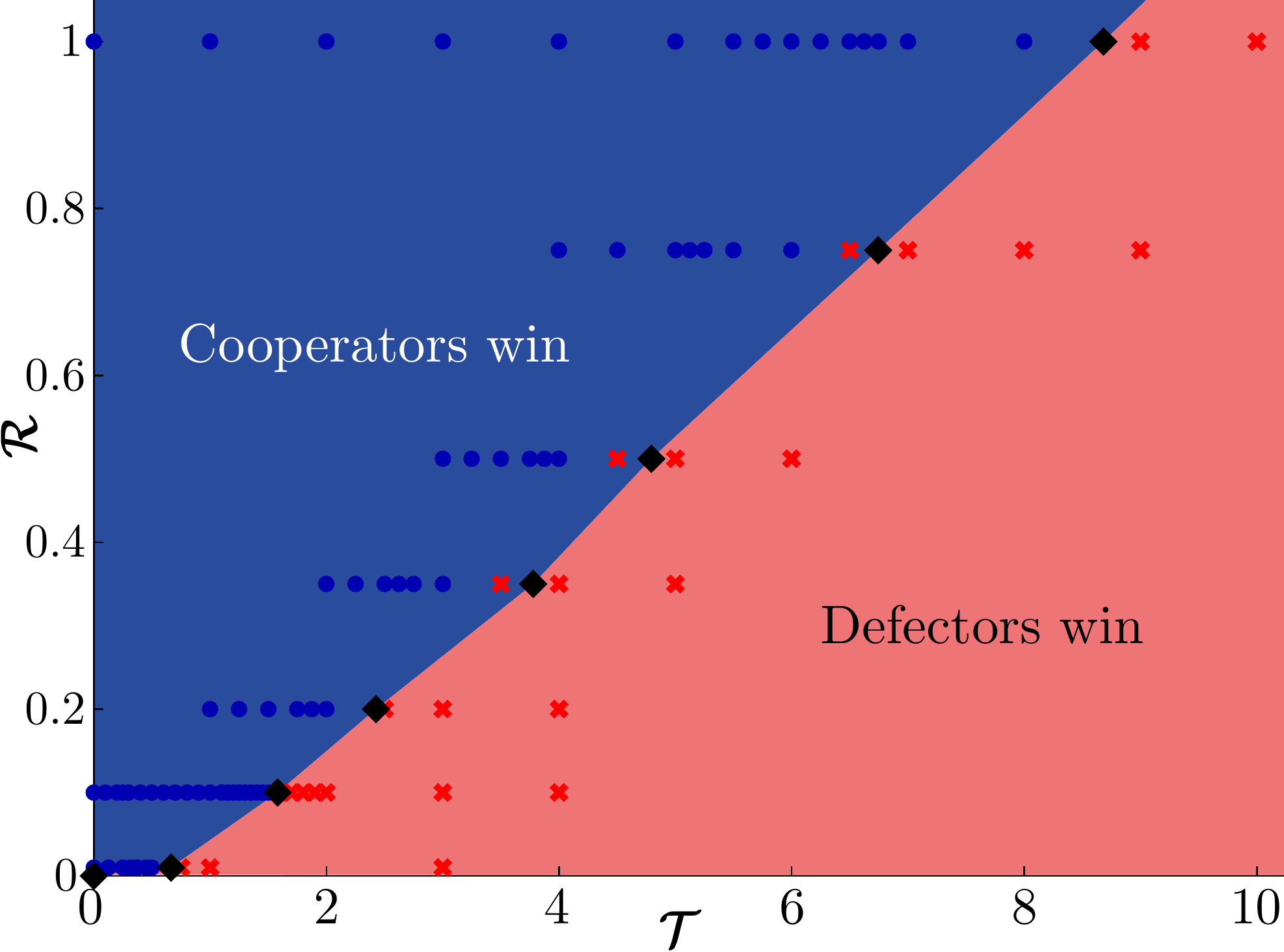}%
%\end{indented}
\caption{\label{fig:phasediagram}%
Phase diagram of the two-species public goods game with range expansion. 
Parameter regimes, where the expanding front fixes either to cooperators or defectors in large systems with $L \to \infty$, are indicated by blue and red shade, respectively. 
Red crosses are from simulations with system size $L = 1000$, where defectors always outcompeted cooperators ($P_\mathrm{fix} = 0$) while blue dots identify events where cooperators survived ($P_\mathrm{fix} > 0$). 
The black diamonds indicate critical points $\mathcal{T}_c$ for $L \to \infty$ determined from finite-size scaling (see Eq.~\eqref{eq:Tcrit} and Fig.~\ref{fig:pfix_C}).
Note that in finite systems defectors have an advantage in a larger parameter region.%
}%
\end{figure}

The above procedure can be repeated for different values of $\mathcal{R}$ to map out the phase diagram (see Fig.~\ref{fig:phasediagram}).
One realizes that the benefit of cooperators from their own species, $\mathcal{R}$, 
has a much more pronounced influence on the final state than the defectors' benefit from cooperating neighbors, $\mathcal{T}\!$. 
This makes sense, since at large times $t \gg 1$ the front contains large single-species sectors. 
Hence, the number of sector boundaries $N_b$, where defectors can benefit from cooperators, is small: $N_b \ll L$.
Therefore, almost all cooperators have cooperating neighbors, while only a few defectors have this advantage.
This is an example of ``preferential assortment'', where the benefit of cooperation is almost entirely
available to other cooperators
%, i.e., related individuals~
\cite{Hamilton:1964wk,Hamilton:1964vp,Griffin:2004kn,Kreft:2004uz,Nowak:1994vj,Ohtsuki:2006cq,Szabo:2007eq,Fu:2010kp}.
So, for a wide range of parameter combinations cooperators can indeed outcompete defectors.
However, for large enough $\mathcal{T}$ the dynamics at the boundaries still determines the final state of the front and defectors outcompete cooperators.

\subsection{Critical exponents of the phase transition \label{sec:crit_exponents}}

In the previous section \ref{sec:phase_trans} we already encountered the critical exponent $\nu_\perp$.
We now continue to determine further critical exponents of the phase transition.
These exponents are universal.
They only depend on the dimension of the system, the number of components of the order parameter, and symmetries 
of the model~\cite{Henkel:2008wn,Odor:2004wm}.
They are independent of microscopic details and do not vary along a phase transition line.
Our system has properties similar to ``compact directed percolation'' (CDP)~\cite{Essam:1989uh}.
This is a stochastic process with a flat front, which also has two distinct absorbing states.
Using this similarity, we proceed by determining critical exponents, which are known for CDP~\cite{Henkel:2008wn}, and compare both models.
\begin{figure}%
\centering%
%\begin{indented}\item[]
\includegraphics[width=\linewidth]{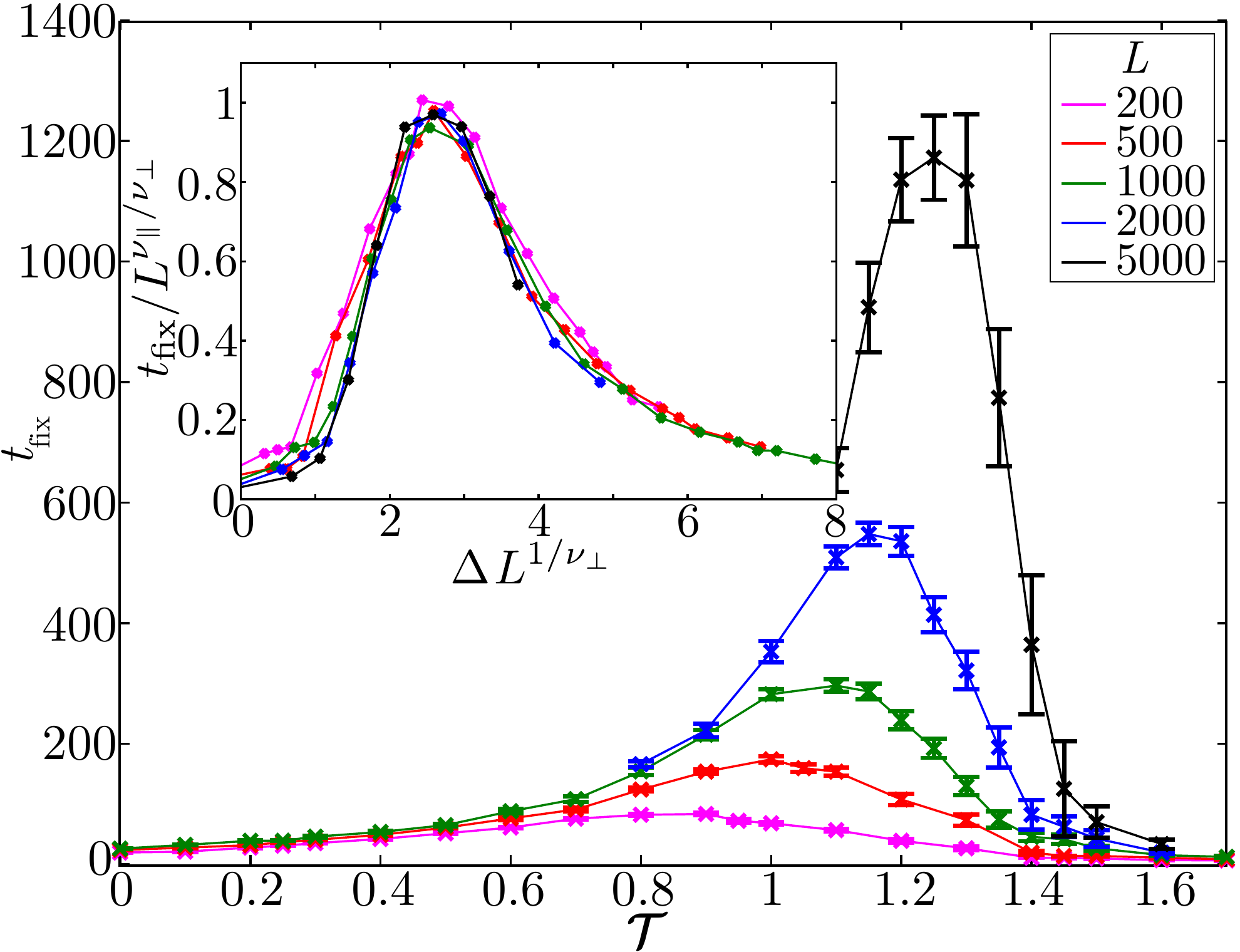}%
%\end{indented}
\caption{\label{fig:tfix}%
Mean time to fixation $t_\mathrm{fix}$ as a function of defector benefit $\mathcal{T}\!$ for various system sizes $L$ 
at $\mathcal{R} = 0.1$.
The inset depicts the rescaled fixation time, where $\Delta = \mathcal{T}_c - \mathcal{T}\!$.
All data collapse on a single master curve for critical exponents $\nu_\parallel = 3.5 \pm 0.1$ and $\nu_\perp = 4.2 \pm 0.1$ and critical point $\mathcal{T}_c =1.59\pm 0.03$. 
Accordingly, $t_\mathrm{fix}$ grows with $L$ and the position of its maximum approaches $\mathcal{T}_c$ for $L \to \infty$.
}
\end{figure}%

At the critical transition, $\mathcal{T} = \mathcal{T}_c$, none of the two species has an advantage.
Heterogeneous fronts, composed of more than one sector, exist for long times before the front fixes to one of the absorbing single-species states.
This can be quantified by the mean time to fixation, $t_\mathrm{fix}$, presented in Fig.~\ref{fig:tfix}.
The data show that the fixation time has a maximum, the position and value of which grow with system size. 

Along the longitudinal direction, in which the front propagates, states are correlated on the longitudinal distance $\xi_\parallel$. 
As before, we expect it to scale like $\left|\Delta\right|^{-\nu_\parallel}$ close to the transition. 
Since the front propagates with a mean velocity, $\xi_\parallel$ is proportional to a correlation time.
Close to $\mathcal{T}_c$ this time becomes very long, which is known as critical slowing down.
Substituting Eq.~\eqref{eq:critscalperp} into
$\xi_\parallel \sim \left|\Delta\right|^{-\nu_\parallel} $,
we find the scaling relation
\begin{align}
\xi_\parallel \sim L^{\nu_\parallel/\nu_\perp} \; .
\end{align}
We expect the mean fixation time to be proportional to
the correlation time 
$ \sim \xi_\parallel$.
Therefore, in the inset of Fig.~\ref{fig:tfix} we plot $t_\mathrm{fix}$ rescaled by $L^{\nu_\parallel/\nu_\perp}$ versus $\Delta$ rescaled by $L^{-1/\nu_\perp}$.
All curves of the main plot collapse on a single master curve for $\mathcal{T}_c = 1.59 \pm 0.03$, and critical exponents $\nu_\perp = 4.2 \pm 0.1$ and $\nu_\parallel = 3.5 \pm 0.1$.
The values of $\mathcal{T}_c$ and $\nu_\perp$ are in good agreement with our fit to Eq.~\eqref{eq:Tcrit}.
So, fixation of the front is determined by the characteristic time
\begin{align}
\tau_\mathrm{fix} \sim \xi_\parallel \sim L^{0.83 \pm 0.05} \; . \label{eq:tfixcrit}
\end{align}

\begin{figure}%
\centering%
%\begin{indented}\item[]
\includegraphics[width=\linewidth]{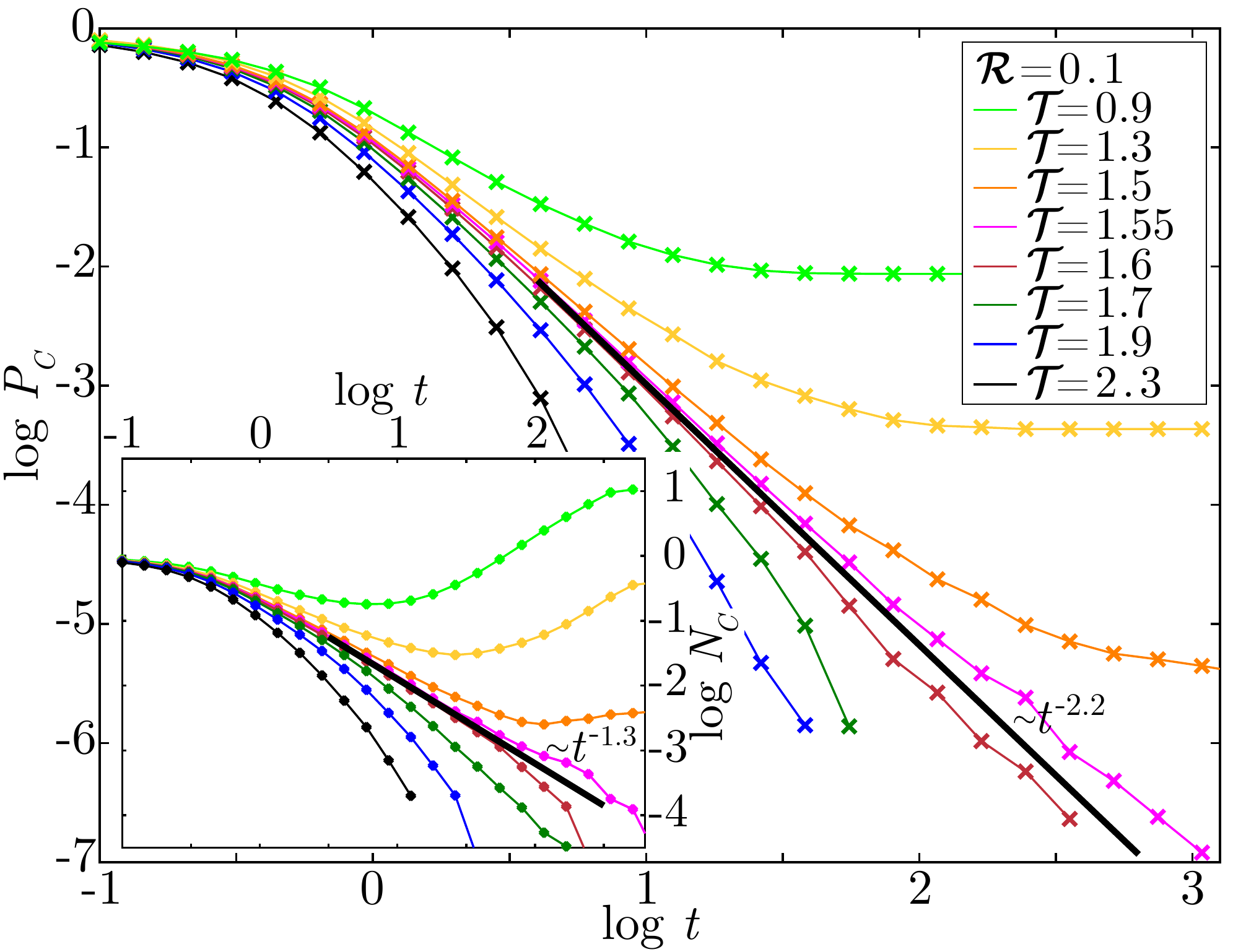}%
%\end{indented}
\caption{\label{fig:single_seed}
Survival probability $P_C$ of cooperators starting from a single site plotted versus time for several values of $\mathcal{T}\!$.
Other parameters are $L = 1000$ and $\mathcal{R} = 0.1$.
For $T > \mathcal{T}_c$, $P_C$ decreases exponentially, while for $T < \mathcal{T}_c$ the front fixes to cooperators with a non-zero probability.
At the transition point $\mathcal{T}_c$, the survival probability decays in time with a power law with exponent $\beta'/\nu_\parallel = 2.2 \pm 0.1$. 
Inset: The number of cooperator sites at the front, $N_C$, decays exponentially in time for $T > \mathcal{T}_c$ and is non-monotonic for $T < \mathcal{T}_c$.
At the transition $N_C$ decreases with a power law with critical exponent $\Theta = 1.3 \pm 0.1$.
}%
\end{figure}

Two more critical exponents right at the transition are related to the survival probability of one species or state, 
which initially occupies a single site while all the other sites are occupied by the other state.
We choose a single cooperator site in a line of defectors and determine the probability $P_C(t)$ that after time $t$ there are still cooperators at the front and also calculate the average number of cooperator sites at the front, $N_C(t)$.
In Fig.~\ref{fig:single_seed} we plot both quantities versus time for different defector benefit $\mathcal{T}$.
At the transition situated between $\mathcal{T} = 1.55$ and 1.6, we find that both $P_C$ and $N_C$ (see inset) decay with power laws in time: 
$P_C(t) \sim t^{-\beta'/\nu_\parallel}$ and $N_C(t) \sim t^{-\Theta}$. 
The respective best fits yield $\beta'/\nu_\parallel = 2.2 \pm 0.1$ and $\Theta = 1.3 \pm 0.1$. 
Using our result for $\nu_\parallel$, we find $\beta' = 7.7 \pm 0.6$.

In general, in phase transitions to absorbing states the critical exponent $\beta$ governs the stationary density of ``active sites'',
when approaching the transition~\cite{Henkel:2008wn}.
In our case, the ``active'' sites can either be cooperators or defectors. 
Since the stationary state is either an all cooperator or an all defector front, the density $ \sim |\mathcal{T}_c - \mathcal{T}|^{\beta}$ jumps from 0 to 1 and, hence, $\beta$ is 0. 
Our results for all the critical exponents are summarized in Table~\ref{tab:critical_exponents}.
\begin{ruledtabular}
\begin{table}%
\centering%
\caption{
Critical exponents for the phase transition to the absorbing states (either long-term global defection or long-term global cooperation)
for the expanding public goods game with frequency dependent selection.
}
\begin{tabular}{ccccc}
$\nu_\perp$ & $\nu_\parallel$ & $\beta$ & $\beta'$ & $\Theta$ \\
\hline
$4.2 \pm 0.1$ & $3.5 \pm 0.1$ & $0$ &$7.7 \pm 0.6$ & $1.3 \pm 0.1$
\end{tabular}
\label{tab:critical_exponents}
\end{table}
\end{ruledtabular}
%
%\begin{table}
%\caption{\label{tab:critical_exponents}
%Critical exponents for the phase transition to the absorbing states (either long-term global defection or long-term global cooperation)
%for the expanding public goods game with frequency dependent selection.
%}
%\begin{indented}
%\item[]
%\begin{tabular}{@{}ccccc}
%\br
%$\nu_\perp$ & $\nu_\parallel$ & $\beta$ & $\beta'$ & $\Theta$ \\
%\mr
%$4.2 \pm 0.1$ & $3.5 \pm 0.1$ & $0$ &$7.7 \pm 0.6$ & $1.3 \pm 0.1$\\
%\br
%\end{tabular}
%%\end{indented}
%\end{table}

The set of critical exponents determines the universality class of a phase transition.
To our knowledge no other non-equilibrium transition has been found, which shares the same set of exponents. 
Hence, the transition between long-term cooperation and long-term defection in our
expanding public goods game with frequency dependent selection constitutes a new universality class.

\subsection{Dynamics of boundaries}

\begin{figure}%
\centering%
%\begin{indented}
%\item[]
\includegraphics[width=\linewidth]{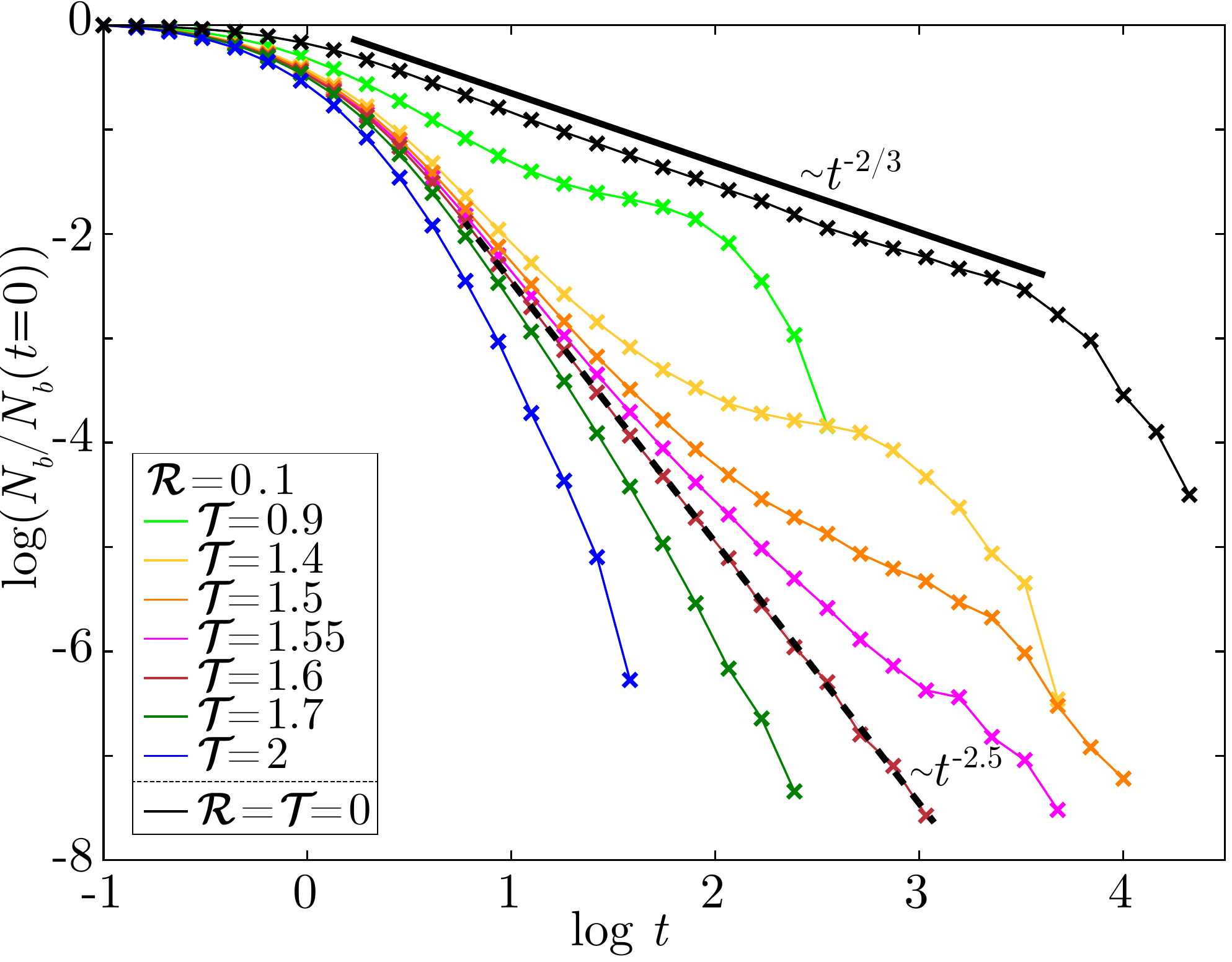}%
%\end{indented}
\caption{\label{fig:nbounds} %
Mean number of sector boundaries $N_b$ plotted as a function of time for several values of $\mathcal{T}\!$.
Other parameters are $L = 1000$ and $\mathcal{R} = 0.1$. 
Solid black line: $N_b(t) \sim t^{-2/3}$, as predicted in Ref.~\cite{Saito:1995wl} for neutral growth ($\mathcal{T} = \mathcal{R} = 0$).
At large times the simulated curve deviates from the power law due to finite system size.
Dashed black line: Close to the critical point $\mathcal{T} = \mathcal{T}_c \approx 1.6$, we find a power-law decay $N_b(t) \sim t^{-2.5}$.
%For neutral growth we recover the literature result~\cite{Saito:1995wl} of $t^{-2/3}$ (bold black line), whereas the behavior 
%is dramatically changed for frequency-dependent selection. In regions where defectors dominate ($\mathcal{T} > 1.6$) 
%the decline of $N_b$ progresses exponentially. Where cooperators are successful ($\mathcal{T} < 1.6$), $N_b$ decreases 
%more slowly, settles temporarily to a plateau, followed by an exponential decay (see main text for details). At the transition 
%between these regimes ($\mathcal{T} = \mathcal{T}_c \approx 1.6$) the number of boundaries decreases as a power law 
%$t^{-5/2}$ (dashed black line).
}
\end{figure}

In this section we investigate the dynamics of the boundaries which separates sectors of cooperators and defectors
from each other. 
In rough fronts the local front orientation is tilted against the main growth direction. When the front grows further,
this tilt directs the movement of boundaries~\cite{Saito:1995wl,Hallatschek:2007gv}. Ultimately, when two of them
meet, they annihilate.
In Fig.~\ref{fig:nbounds} we plot their mean number $N_b$ versus time for several values of the
defector benefit $\mathcal{T}$. Right at the transition (dashed black line), $N_b$ shows a  power law decay. We now 
discuss the different regimes in Fig.~\ref{fig:nbounds}.

For neutral systems, where frequency-dependent selection is absent ($\mathcal{T} = \mathcal{R} = 0$), 
any inclination of the front is created by stochastic surface or Eden 
roughening~\cite{Barabasi:1995vz,Krug:1992wu,HalpinHealy:1995wb,Odor:2004wm}.
The surface undulations obey KPZ-scaling~\cite{Kardar:1986vl}
and thereby drive the decay of $N_b$~\cite{Saito:1995wl,Hallatschek:2007gv}.
Boundaries move superdiffusively along the front with a mean-square displacement proportional to $t^{4/3}$~\cite{Derrida:1991tv}.
On average, they annihilate after having traveled the mean distance $L/N_b$ between the boundaries, for which they need the time $ \sim (L/N_b)^{3/2}$. 
Hence, boundaries annihilate with a rate proportional to $N_b^{3/2}$, which implies
\begin{align}
\dot N_b \sim - N_b^{3/2} N_b \; . \label{eq:Nbdot}
\end{align}
%
% for $t \gg 1$ it follows that 
So, the number of boundaries decreases as
\begin{align}
N_b(t) \sim t^{-2/3} \; , \label{eq:Nboft}
\end{align}
as already observed by Saito and M\"uller-Krumb\-haar~\cite{Saito:1995wl}.
This power law is excellently reproduced by our simulations in the case of neutral growth, $\mathcal{R} = \mathcal{T} = 0$,
as illustrated by the solid black line in Fig.~\ref{fig:nbounds}.

For $\mathcal{R} \ne 0 $ and $\mathcal{T} \ne 0$, species reproduce with different rates. Hence, the fronts
of two neighboring sectors (occupied by different species) advance with different speeds. This influences the tilt of the front 
orientation, in addition to stochastic roughening in neutral systems, and thereby the movement of the separating boundary.  
Thus, we do not expect Eq.~\eqref{eq:Nbdot} to be valid.
%A boundary's position determines the local expansion speeds of the front and consequently the orientation of the front at that position.
%As the movement of boundaries is in part controlled by the orientation of the front, a feedback is created, which induces non-trivial %correlations between front position and boundaries.
%This prohibits an approach similar to that of Eq.~\eqref{eq:Nbdot}.}

Indeed, Fig.~\ref{fig:nbounds} reveals different regimes for the mean number of boundaries $N_b$.
For $\mathcal{T} > \mathcal{T}_c$, $N_b$ decays exponentially in time in line with the exponential decay of the survival probability $P_C$ in Fig.~\ref{fig:single_seed} 
and similar to the case of selective advantage in Ref.~\cite{Saito:1995wl}. 
For $\mathcal{T} < \mathcal{T}_c$ boundaries annihilate less frequently.
Narrow defector sectors persist in the front dominated by cooperators since almost all individuals in the defector sectors have cooperating neighbors.
This results in the upward curvature of the curves in Fig.~\ref{fig:nbounds}.
However, the defector sectors cannot expand laterally and ultimately loose contact to the front due to random fluctuations, and $N_b$ declines exponentially.
At $\mathcal{T}_c$ the number of boundaries decreases with a power law $N_b \sim t^{-\chi}$, where a new exponent $\chi \approx 2.50 \pm 0.05$ appears.
This power law implies that 
%by coarsening 
the number of boundaries declines from the initial value $N_b(t=0) \sim L$ to the order of 1 in the coarsening time
\begin{align}
\tau_\mathrm{coarse} \sim L^{1/\chi} \approx L^{0.40 \pm 0.01} \; . \label{eq:tcoarse}
\end{align}
Comparing with Eq.~\eqref{eq:tfixcrit} reveals $1/\chi < \nu_\parallel/\nu_\perp$.
This suggests that for large systems fixing the front to one species takes much longer than coarsening to a few sectors.
Hence, the few remaining boundaries move 
%qualitatively 
differently compared to early times since they have to annihilate to fix the front to a single species.

%For the classical Eden model any inclination of the front is created by stochastic surface roughening~\cite{Barabasi:1995vz,Krug:%1992wu,HalpinHealy:1995wb,Odor:2004wm}.
%Since the scaling properties of the roughness 
%(i.e.\ the typical transverse and longitudinal extensions of surface undulations and the corresponding time scales on which they exists)
%are known in this case, it was possible to derive how boundaries move transversely during expansion~\cite{Saito:%1995wl,Hallatschek:2007gv}.
%In our case this ``Eden roughening'' is not the only source of undulations in the front:
%As a result of different reproduction rates, species advance with different speeds, and an additional source of ``selective roughness'' is encountered, c.f.\ Section \ref{sec:phen}.
%This influences the boundary dynamics, and might be the reason to the new scaling regime of $N_b$ encountered in Fig.~
%\ref{fig:nbounds}.

\begin{figure}%
\centering%
%\begin{indented}
%\item[]
\includegraphics[width=\linewidth]{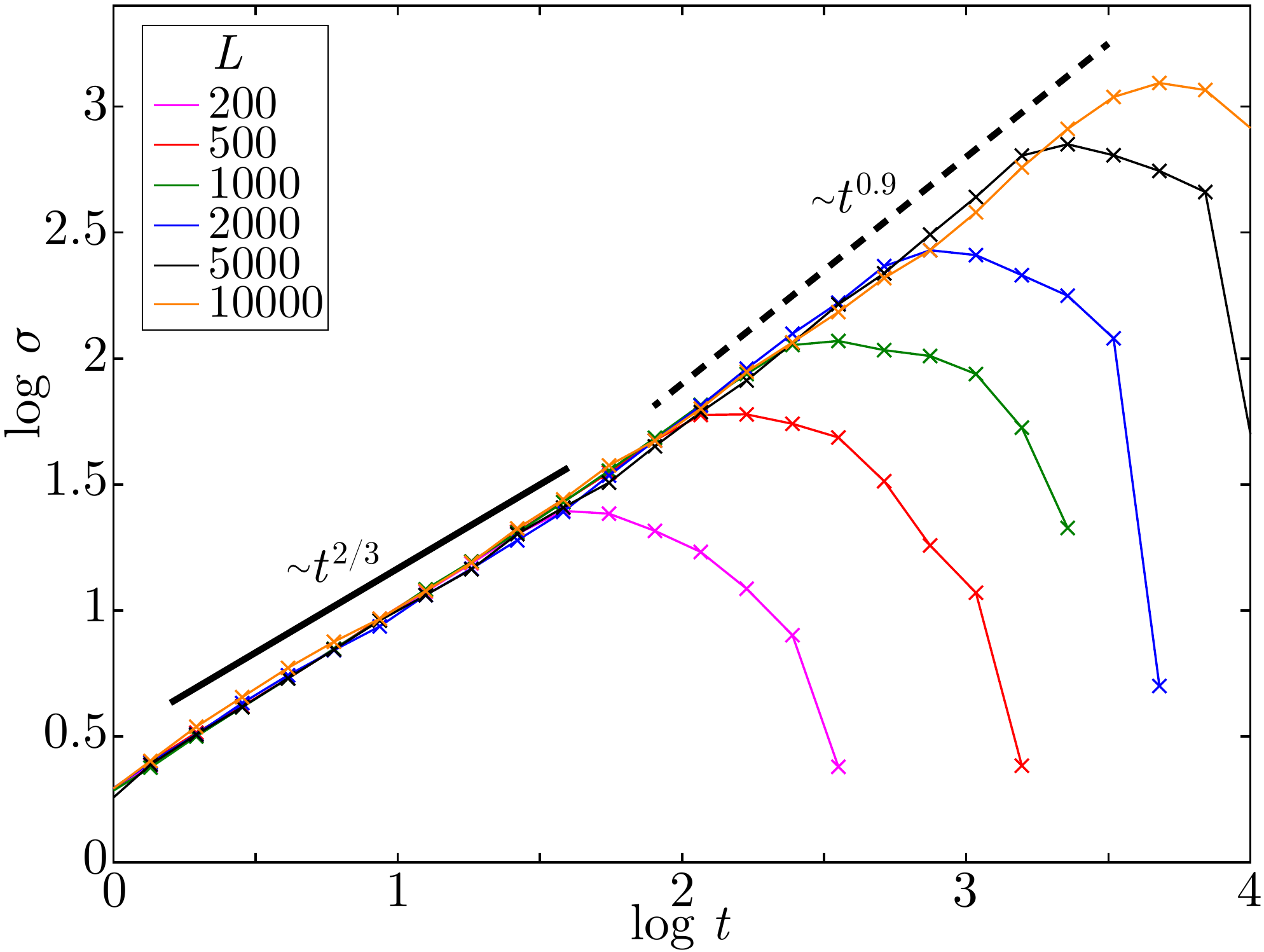}%
%\end{indented}
\caption{\label{fig:sigmaboundscaled}
Time evolution of two boundaries with initial distance $L/2$ in a system close to criticality ($\mathcal{T} = 1.6$) and
at $\mathcal{R} = 0.1$.
Standard deviation $\sigma$ of the transverse distance plotted versus time $t$ for several system sizes $L$. 
The boundaries move superdiffusively. Initially, $\sigma \sim t^\eta$ with $\eta = 0.71 \pm 0.05$, and consistent with  
Eden scaling ($t^{2/3}$, bold black line).
At later times a crossover to $\sigma \sim t^{\eta'}$ with $\eta' = 0.9 \pm 0.1$ occurs (dashed black line).
%This indicates that different reproduction rates dominate large-scale surface undulations, which in turn influence the boundaries' movements.
}
\end{figure}%

To check if this is the case, we employ the initial condition, where the front is composed of only two sectors of size $L/2$ each, separated by two boundaries.
We quantify the boundaries' random motion by monitoring the temporal evolution of the standard deviation for the transverse distance $\ell_\perp$,
\begin{align}
\sigma(t) := \sqrt{\langle [\ell_\perp(t) - \langle\ell_\perp(t)\rangle]^2 \rangle} \, . \label{eq:stddev}
\end{align}
We subtract the mean distance $\langle\ell_\perp(t)\rangle$ to take care of any transient drift, when the front relaxes
from its initially flat to the rough shape, and an expected small drift if $\mathcal{T}$ is not exactly $\mathcal{T}_c$.

From Fig.~\ref{fig:sigmaboundscaled} we see that, for early times, $\sigma$ grows like a power law, $\sigma \sim t^\eta$, 
with $\eta = 0.71 \pm 0.05$. This is consistent with meandering 
boundaries induced by Eden roughening, $ \sigma \sim t^{2/3}$~\cite{Derrida:1991tv,Saito:1995wl,Hallatschek:2007gv}. 
We expect such a behavior since the roughness of the front has not fully developed yet.
For large systems and later times we find a crossover to $\sigma \sim t^{\eta'}$ with $\eta' = 0.9 \pm 0.1$.
%This is close to ballistic scaling, $\sim t^1$, which has been conjectured for boundaries located at indentations between two sectors which expand with identical speeds~\cite{Derrida:1991tv}.
This confirms our earlier statement that at late times the few  boundaries remaining after coarsening move differently. Indeed,
they show an even stronger superdiffusive motion than Eden scaling, which is associated with the long-lived and pronounced 
surface undulations in systems with frequency-dependent selection.

\subsection{Surface roughening of the expanding front}
%\jtk{In the previous section we argued that long-scale surface undulations regulate boundary movements.}
%How are these undulations different from those observed in the classical Eden model? 
\begin{figure}%
\centering%
%\begin{indented}
%\item[]
\includegraphics[width=\linewidth]{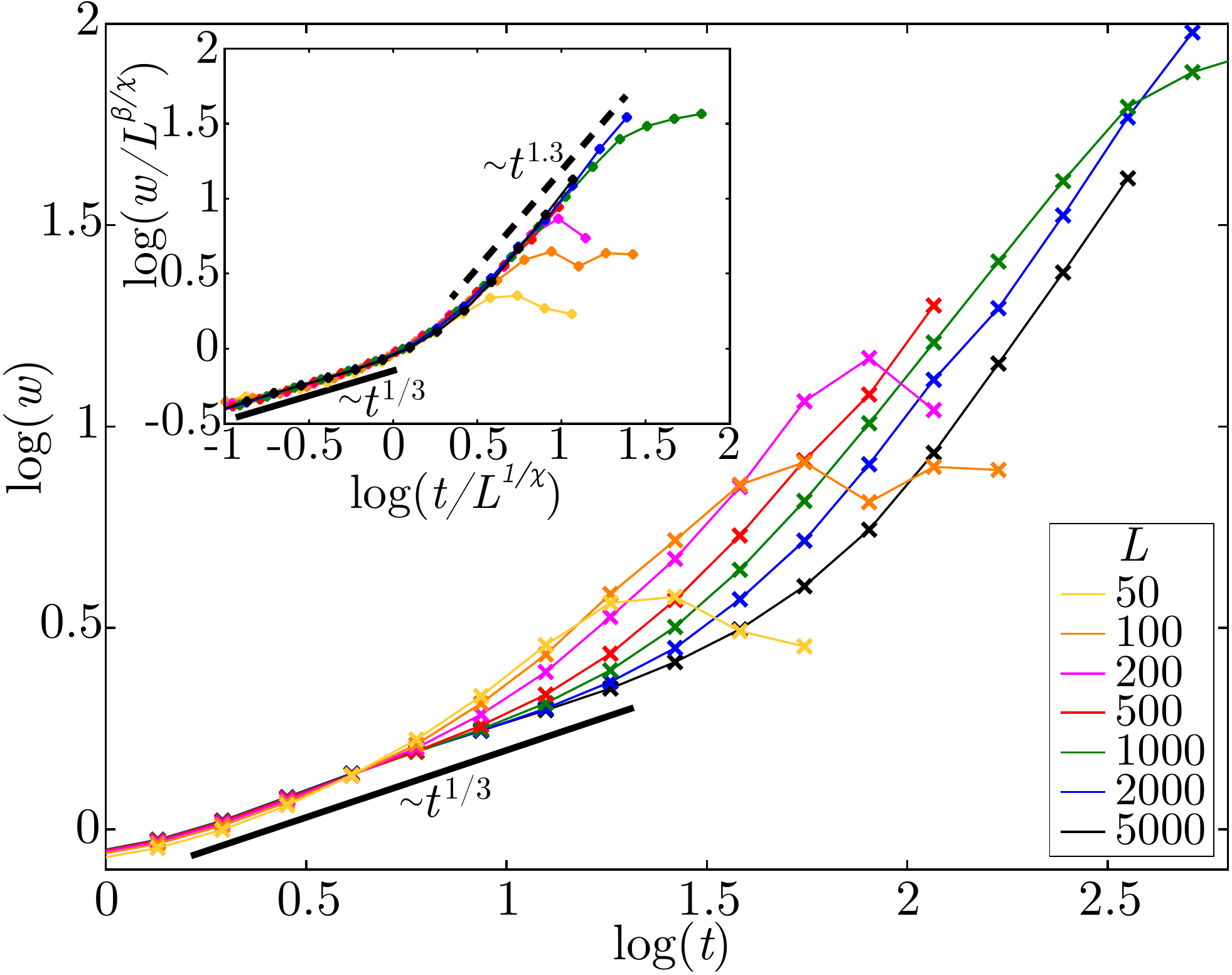}%
%\end{indented}
\caption{\label{fig:widthvst}
The width $w$ of the front plotted versus time for several system sizes $L$ close to the critical point at
$\mathcal{T} = \mathcal{T}_c \approx 1.6$.
Inset: After rescaling time $t$ by $\tau_\times \sim L^{1/\chi}$ and width $w$ by $w_\times \sim L^{\gamma/\chi}$, the curves
collapse onto a single master curve. 
The solid black line indicates Eden roughening with $w \sim t^{1/3}$, whereas the
dashed black line shows selective roughening with $w \sim t^{1.3}$ for $t \gg \tau_\times$.
Note that the saturation of $w$ at large times is due to the finite system size.
}
\end{figure}%

We now discuss the surface roughness or undulations of the expanding front and compare our results to the classical Eden model.
We measure the roughness of a front, which has not yet fixed to one species, by calculating its width $w$ for system size $L$ and at time $t$:
\begin{align}
w(L,t) := 
\bigg( \frac1L \sum_{i=1}^L\left[h(i,t) - \bar h(t)\right]^2 \bigg)^{1/2}\; \!\!\!. \label{eq:def_width}
\end{align}
Here, $h(i,t)$ is the longitudinal position of the front at its transverse site $i$ and
%the local position of the front.
%From our simulations, we take $h(i,t)$ to be the most advanced occupied site at transverse position $i$, i.e.\ the largest %longitudinal position $j$ for which $s_{i,j} \neq 0$.
$\bar h(t)$ is the mean position
\begin{align}
\bar h(t) &:= \frac1L\sum_{i=1}^L h(i,t)\; .
\end{align}
Figure~\ref{fig:widthvst} plots the width $w(L,t)$ close to the critical point at $\mathcal{T} = \mathcal{T}_c \approx 1.6$.
Initially, the roughness of the front grows like in the original Eden model~\cite{Eden:1960vd}: $w \sim t^\gamma$, 
where the growth exponent $\gamma = 1/3$ belongs to the KPZ-universality class~\cite{Barabasi:1995vz}. 
At intermediate times a new regime sets in, where $w \sim t^{\gamma'}$ increases with enhanced growth exponent $\gamma' \approx 1.3 \pm 0.1$. 
This marks the transition from Eden roughening to ``selective roughening''. 
Here, the typical shape of the front is determined by
%marked by advanced 
advancing cooperator sectors and trailing defector sectors and ultimately drives the accelerated increase in the front's width $w$.
The crossover to this regime happens at time $\tau_\times$, which increases with system size $L$ as Fig.~\ref{fig:widthvst} shows.
This makes sense since we expect selective roughening to dominate over Eden roughening when the lateral extension of sectors is comparable to $L$,
%Eden roughening happens at all length scales up to $L$ and therefore contributes more to roughness for larger systems, 
%whereas selective roughening can only contribute on length scales up to the typical extension of sectors.
i.e., $\tau_\times \sim \tau_\mathrm{coarse} \sim L^{1/\chi}$.
Due to Eden roughening the width at the crossover is $ w_\times \sim \tau_\times^\gamma \sim L^{\gamma/\chi} \approx L^{0.13}$. 
Indeed, rescaling width and time with $w_\times$ and $\tau_\times$, respectively, 
collapses all data in Fig.~\ref{fig:widthvst} onto a single master curve, as the inset demonstrates.
To conclude, surface roughening close to criticality occurs in two regimes. 
Until crossover time $\tau_\times$, one observes Eden roughening, 
whereas for times larger $\tau_\times$ selective roughening occurs until the front fixes to one species.
The dynamics of the width of the front is summarized by 
\begin{align}
w(L,t) \sim
\begin{cases}
t^\gamma & t \ll \tau_{\times} \sim L^{1/\chi} \; ,\\
t^{\gamma'} L^{(\gamma-\gamma')/\chi}& t \gg  \tau_{\times}\; . 
%\tau_\mathrm{sat}  \gg t \gg \tau_{\times}\\
%L^{1.2}& t \gg \tau_\mathrm{sat}  \sim L^{1.2}
\end{cases}
\end{align}
%
%\begin{eqnarray}
%w(L,t) \sim
%\left\{
%\begin{array}{ll}
%t^\gamma & t \ll \tau_{\times} \sim L^{1/\chi} \; ,\\
%t^{\gamma'} L^{(\gamma-\gamma')/\chi}& t \gg  \tau_{\times}\; .
%\end{array}
%\right.
%\end{eqnarray}
%This result again emphasizes the qualitative change of the system's dynamics as extended sectors are created. 
%As cooperator sectors advance and defectors fall back, the front's width widens. 

\section{Summary and Conclusion\label{sec:discussion}}

In this work we studied a generalized Eden model, where two species compete with each other at the rough expanding front.
Individuals of the two species influence each other by frequency-dependent selection, which acts between nearest neighbors.
We analyzed the evolutionary dynamics at the expanding front, where single-species sectors form and coarsen.
Ultimately, the front fixes to one species, which we identify with an absorbing state of our model.

In its general form the model can implement several scenarios including selective advantage, 
and also well-known game theoretical settings like the snow drift game or the coordination game.
Each of them creates distinct patterns, which should be analyzed in detail in future work.
% and compared to experiments. 
For the prominent example of a public goods game, we find that cooperators prevail in a wide parameter regime, as expected for a spatial version of a social dilemma~\cite{Nowak:1994vj,Ohtsuki:2006cq,Szabo:2007eq,Fu:2010kp}.
For other parameter values defectors take over the front, as usual.

We identify the transition between long-term cooperation and long-term defection as a nonequilibrium critical phase transition between two absorbing states.
The set of critical exponents (see Table~\ref{tab:critical_exponents}), which we determined by analyzing critical and 
finite size scaling, shows that the phase transition belongs to a new universality class. 
We attribute this result to the fact that the front in our model is rough and not flat as in usual absorbing states.
Close to the critical transition the front's roughness exhibits a crossover in time from slow Eden roughening to fast selective roughening.
Strong roughening has also been observed at 
%roughness has been associated with 
phase transitions in a related model by Lavrentovich and Nelson~\cite{Lavrentovich:2014jo}.

Interestingly, the critical exponents determined in our present work violate the so-called generalized hyperscaling relation~\cite{Mendes:1994wj}
\begin{align}
d \nu_\perp = \nu_\parallel \Theta + \beta +\beta' \, ,
\end{align}
where the number of transverse dimensions $d$ is 1. %in our case. 
This relation holds for most universality classes of phase transitions to absorbing states~\cite{Henkel:2008wn}. 
It is not obvious why our 
%system is an exception.
model does not obey the scaling relation.

The roughness of the front correlates with superdiffusive motion of the boundaries separating sectors.
Two factors contribute to the movement of the boundaries on long length scales.
On the one hand, the direct competition between the species on either side of the boundaries pushes them towards the sector composed of the more slowly reproducing species.
On the other hand, boundaries follow the local tilt of the front.
In the public goods game cooperator sectors are advanced, while defector sectors lag behind.
Near the critical transition, defectors outcompete their direct cooperating neighbors but the front is tilted towards 
sectors filled by defectors, so the two factors move the boundaries in opposite directions.
At the phase transition both effects cancel and the front fixes with equal probability to either species. 
The strong roughening correlates with superdiffusive motion of the boundaries with nearly ballistic scaling.

Accordingly, whether a species takes over the expanding front is determined by two contributions: 
its reproduction rate relative to its competitor and its position relative to the average front position. 
The influence of different reproduction rates of neighboring species
can directly be compared and is summarized in the phrase ``survival of the fittest''.
The position at the front determines the available space for progeny,
%(the ``founder effect''~\cite{Edmonds:2004il,Travis:2007iy}), 
which then have the opportunity to expand sidewards.
This is illustrated by the phrase ``survival of the fastest''~\cite{VanDyken:2013tf}. 
%Where these two effects balance, there exists a transition between cooperator vs.\ defector dominance.

%		Outlook
In our model the number of sectors only decreases.
It does not include experiments with mutually beneficial interactions between different species, which do not generate sectors~\cite{Muller:2014ev,Momeni:2013ee,Kovacs:2014ev}.
In future extensions of our model this may be remedied by including motility of individuals~\cite{Reichenbach:2007hk,Gelimson:2013da}, 
by allowing reproduction to more distant lattice sites~\cite{Lavrentovich:2014jo}, or by increasing the maximal number of individuals per lattice site from one.
%(i.e.\ ``deme size'' larger 1)~\cite{MotooKimura:1964wy,Korolev:2011jk}. 
Moreover, it is worthwhile to consider interactions ranging beyond nearest neighbors,
%longer ranged interactions, 
since biomolecules, released by individual microorganisms, may diffuse in the extracellular medium~\cite{Julou:td,Allen:2013hl,Menon:2015wm}. 
In the public goods game scenario this would stabilize narrow sectors of defectors so that they do not lose contact to the front.

%Fin
In general, range expansion of multiple species will develop enhanced roughness at the growing front.
As we demonstrated here, the corresponding models have new and interesting statistical properties.
From a biological point of view, roughness is important. 
It affects the territories that different species occupy and thereby their evolutionary success through the strong random motion of sector boundaries.
This may also be relevant for range expansion in a real environment and not just in a test tube.
To better understand the properties and consequences of rough expanding fronts, further theoretical work is needed.
At the same time further experiments should look for the fingerprint of roughness in microbial colony growth.

\begin{acknowledgments}
We thank the research training group GRK 1558 funded by Deutsche Forschungsgemeinschaft for financial support.
We further thank Erwin Frey, Maria Eckl, Florian Gartner, and Raphaela Ge\ss{}ele for discussion and collaboration on a related model.
\end{acknowledgments}

\bibliography{generaltwospecieseden}

\end{document}